\documentclass[a4paper,twocolumn]{scrartcl}

\title{The right tool for the right question --- beyond the encoding versus decoding dichotomy}
\author{Sebastian Weichwald, Moritz Grosse-Wentrup \\
    {\small Max Planck Institute for Intelligent Systems, T{\"u}bingen, Germany}\\
    {\small\texttt{[sweichwald, moritzgw]@tue.mpg.de}}
}
\date{}

\usepackage[square]{natbib}
\usepackage{pgfplots}
\usepackage{amsmath}
\usepackage{subcaption}
\usepackage{xfrac}
\renewcommand{\frac}[2]{\sfrac{#1}{#2}}
\DeclareMathOperator{\cov}{cov}
\DeclareMathOperator{\var}{var}

\usepackage[hidelinks]{hyperref}

\usepackage{pgfplots}
\usepgfplotslibrary{colormaps}
\pgfplotsset{compat=1.12}
\tikzset{ellC/.style={/utils/exec={\pgfplotscolormapdefinemappedcolor{#1}}, draw=mapped color, fill=mapped color, text=white}}
\tikzset{observed/.style={rectangle, minimum size=1cm}}
\tikzset{unobserved/.style={draw=gray, text=gray, observed}}
\usetikzlibrary{shapes}
\usepackage{ifthen}

\newcommand{\brainS}[7]{%
\ifthenelse{\equal{#4}{}}{%
    \node[unobserved] at (#1+1,#2+2) {$V$};
    }{%
    \node[ellC=#4, observed] at (#1+1,#2+2) {$V$};
}
\ifthenelse{\equal{#5}{}}{%
    \node[unobserved] at (#1+3,#2+2) {$D$};
    }{%
    \node[ellC=#5, observed] at (#1+3,#2+2) {$D$};
}
\ifthenelse{\equal{#6}{}}{%
    \node[unobserved] at (#1,#2+0) {$C$};
    }{%
    \node[ellC=#6, observed] at (#1,#2+0) {$C$};
}
\ifthenelse{\equal{#7}{}}{%
    \node[unobserved] at (#1+2,#2+0) {$T$};
    }{%
    \node[ellC=#7, observed] at (#1+2,#2+0) {$T$};
}
\node[] at (#1+1,#2+-2) {#3};
}

\newcommand{\brainR}[7]{%
\ifthenelse{\equal{#4}{}}{%
    \node[unobserved] at (#1,#2+2) {$T$};
    }{%
    \node[ellC=#4, observed] at (#1,#2+2) {$T$};
}
\ifthenelse{\equal{#5}{}}{%
    \node[unobserved] at (#1+2,#2+2) {$C$};
    }{%
    \node[ellC=#5, observed] at (#1+2,#2+2) {$C$};
}
\ifthenelse{\equal{#6}{}}{%
    \node[unobserved] at (#1+1,#2+0) {$M$};
    }{%
    \node[ellC=#6, observed] at (#1+1,#2+0) {$M$};
}
\ifthenelse{\equal{#7}{}}{%
    \node[unobserved] at (#1+3,#2+0) {$D$};
    }{%
    \node[ellC=#7, observed] at (#1+3,#2+0) {$D$};
}
\node[] at (#1+2,#2+-2) {#3};
}

\pgfplotsset{colormap={divcolor}{%
rgb (0cm)= (0.230, 0.299, 0.754)%
rgb (1cm)= (0.915, 0.915, 0.915)
rgb (2cm)= (0.085, 0.532, 0.201)%
}}

\begin{document}

\maketitle

There are two major questions that neuroimaging studies attempt to answer:
First, how are sensory stimuli represented in the brain (which we subsequently term the \textit{stimulus-based} setting)?
And, second, how does the brain generate cognition (subsequently termed the \textit{response-based} setting)?
There has been a lively debate in the neuroimaging community whether encoding and decoding models can provide insights into these questions (e.\,g.~\citep{naselaris2011encoding,todd2013confounds,davis2014differences,haufe2014interpretation,woolgar2014coping,weichwald2015causal}).
In this commentary, we construct two simple and analytically tractable examples to demonstrate that while an encoding model analysis helps with the former, neither model is appropriate to satisfactorily answer the latter question.
Consequently, we argue that if we want to understand how the brain generates cognition, we need to move beyond the encoding versus decoding dichotomy and instead discuss and develop tools that are specifically tailored to our endeavour.

Since the disagreement among researchers can partially be attributed to differing use of terminology, we begin by briefly introducing the two (types of) models that are subject of the ongoing debate and make explicit the terminology used throughout this comment.
On the one hand, there are (univariate) \emph{encoding models} (also referred to as forward or generative models) that approximate each neurophysiological variable $X_i$ as a function of the experimental condition $Y$ (e.g.~statistical parametric mapping~\citep{friston1994statistical}), i.\,e.
\[
    \begin{bmatrix} \widehat{X}_1 \\ \vdots \\ \widehat{X}_n \end{bmatrix}
    =
    \operatorname{enc}(Y)
    =
    \begin{bmatrix} \operatorname{enc}_1(Y) \\ \vdots \\ \operatorname{enc}_n(Y) \end{bmatrix}.
\]
On the other hand, there are (multivariate) \emph{decoding models} (also referred to as backward or discriminative models or multi-voxel pattern analysis (MVPA)) that predict the experimental condition $Y$ given the neurophysiological variables $X_1, \ldots, X_n$~\citep{mitchell2004learning,pereira2009machine}, i.\,e.
\[
    \widehat{Y} = \operatorname{dec}(X_1,\ldots,X_n).
\]

It has been argued that encoding models allow for richer interpretations than decoding models and can in principle provide a complete functional description of a brain state variable~\citep{naselaris2011encoding,haufe2014interpretation}.
Decoding models are often considered inherently difficult to interpret and may determine neurophysiological variables as relevant that are statistically independent of the experimental condition~\citep{todd2013confounds,woolgar2014coping,haufe2014interpretation}.
Indeed, the difference between the two models in terms of their interpretation has experimentally been confirmed~\citep{huth2016decoding,bach2017reward}.
Further distinguishing between stimulus- and response-based paradigms has enabled a comprehensive overview over the theoretical limitations of interpreting either model alone or both at the same time~\citep{weichwald2014causal,weichwald2015causal}.

We contribute to this discussion by providing two instructive examples that illustrate the fundamental problems in the interpretation of encoding and decoding models.
Specifically, we show that even in the infinite-sample limit and under correct model assumptions the exclusion of relevant variables---e.\,g.\ when preselecting regions of interest or because of unobserved (latent) brain processes---may lead to incorrect conclusions about the qualitative and quantitative relations of brain processes and experimental variables.
Our examples underline the importance of developing methods that are robust against such confounds.

\section{Examples}

In the following, we construct two hypothetical ground-truth models, the first for a stimulus- and the second for a response-based paradigm, and consider what interpretations one would obtain from encoding and decoding models fit to differing subsets of variables.
This enables us to study deviations between true and estimated relations of brain processes and experimental variables under idealistic conditions.%
\footnote{First, the assumptions for ordinary least squares regression are met since we assume the ground-truth models to be linear and Gaussian. Second, we consider the true ordinary least squares weights one would obtain in the limit of infinite data, hence ignoring finite sample estimation errors~\citep{taylor2015statistical}.}

\subsection{Stimulus-based paradigm}\label{secStimExample}

Assume we conduct a stimulus-based experiment to investigate the effect of a stimulus variable $S$ onto neurophysiological variables (e.\,g.\ showing images of different brightness to subjects while recording their brain activity in different regions of interest).
Assume further that the true relationships are governed by the additive Gaussian noise model described in Figure~\ref{figStimulus} where upper and lower case letters denote variables and their linear relations, respectively.
The linear effect of $S$ onto each variable is determined by the sum of weight products along each path from $S$ to that variable;
e.\,g.\ the effect of $S$ onto $D$ is linear with slope $ab + ade = -2$ (i.\,e.\ increased image brightness leads to a twice as much decreased average activity in $D$\footnote{%
More precisely, if we were to intervene and forcefully set $S$ to take the value $5$ then $D$ will follow a Gaussian distribution with mean $-10$ while in the observational setting the mean is $0$.
It is these differences in the distribution of a variable (the effect) upon an intervention on another variable (the cause) that form the basis of the conceptualisation of interventional causation in structural equation models~\citep{spirtes2000causation,pearl2009causality}}).
The true effects of $S$ onto $V,D,C,T$ are $1,-2,1,2$ respectively.

In Figure~\ref{figStimulusPlots} we show toy brain maps depicting the true effects of $S$ onto each variable as well as the outcome one would obtain when running one of the following three analyses of a system admitting the above ground-truth model:
\begin{description}

\item[Analysis 1: $\operatorname{enc}(S)$]\ \\
A mass-univariate encoding analysis where for each variable $X \in \{V,D,C,T\}$ we interpret the weight $\beta_X$ in the model
$$\widehat{X} = \operatorname{enc}(S) = \alpha_X + \beta_X S$$
to reflect the effect of $S$ onto $X$.
The true ordinary least squares weights one would obtain in the limit of infinite data are
\begin{align*}
\begin{bmatrix}
    \beta_V \\ \beta_D \\ \beta_C \\ \beta_T
\end{bmatrix}
=
\frac{1}{\var(S)}
\begin{bmatrix}
    \cov(V,S) \\ \cov(D,S) \\ \cov(C,S) \\ \cov(T,S)
\end{bmatrix}
=
\begin{bmatrix}
    1 \\ -2 \\ 1 \\ 2
\end{bmatrix}
\end{align*}

\item[Analysis 2: $\operatorname{dec}(V,D,C,T)$]\ \\
A decoding analysis including the variables $V,D,C,T$ where we interpret the weights of the model
$$\widehat{S} = \alpha + \beta V + \gamma D + \delta C + \epsilon T$$
to reflect the relation between $S$ and each respective variable.
The true ordinary least squares weights one would obtain in the limit of infinite data are
\begin{align*}
\begin{bmatrix}
    \beta \\ \gamma \\ \delta \\ \epsilon
\end{bmatrix}
=
\Sigma_{V,D,C,T}^{-1}
\begin{bmatrix}
    \cov(V,S) \\ \cov(D,S) \\ \cov(C,S) \\ \cov(T,S)
\end{bmatrix}
=
\begin{bmatrix}
    \frac{1}{2} \\ 0 \\ 0 \\ 0
\end{bmatrix}
\end{align*}

\item[Analysis 3: $\operatorname{dec}(D,C,T)$]\ \\
A decoding analysis as above, this time excluding the variable $V$.
The true ordinary least squares weights one would obtain in the limit of infinite data are
\begin{align*}
\begin{bmatrix}
    \gamma \\ \delta \\ \epsilon
\end{bmatrix}
=
\Sigma_{D,C,T}^{-1}
\begin{bmatrix}
    \cov(D,S) \\ \cov(C,S) \\ \cov(T,S)
\end{bmatrix}
=
\begin{bmatrix}
    \frac{-1}{7} \\ \frac{-1}{7} \\ \frac{1}{7}
\end{bmatrix}
\end{align*}

\end{description}

In our scenario, the encoding analysis would indeed reveal the correct effects of $S$.
If instead we were to interpret the weights of the full linear decoding model $\operatorname{dec}(V,D,C,T)$ we would arrive at incorrect interpretations; e.\,g.\ in this model $\delta$ turns out to be zero suggesting that $S$ is not related to $C$.
Further complicating the matter, if $V$ was excluded from the analysis (due to subjective variable selection criteria or being unobserved), i.\,e. when interpreting $\operatorname{dec}(D,C,T)$, we would arrive at yet other interpretations; e.\,g.\ we would obtain $\delta = -\frac{1}{7}$, which may mislead researchers to believe that $S$ has an inhibitory effect on $C$.

\begin{figure*}
\begin{minipage}{.5\textwidth}
\begin{tikzpicture}
    \node (T) at(0, -3) {$T${\tiny emporal}};
    \node (C) at(3, -3) {$C${\tiny entral inferotemporal}};
    \node (V) at(1.5, -1.5) {$V${\tiny isual}};
    \node (S) at(1.5, 0) {$S${\tiny timulus}};
    \node (D) at(4.5, -1.5) {$D${\tiny isturbance}};
    \node (H) at(1.5, -4.5) {$H${\tiny idden}};
    \draw[<-] (T) -- node[auto] {$c = 2$} (V);
    \draw[<-] (C) -- node[auto] {$d = 1$} (V);
    \draw[<-] (V) -- node[auto] {$a = 1$} (S);
    \draw[->] (C) -- node[right] {$e = -1$} (D);
    \draw[->] (V) -- node[auto] {$b = -1$} (D);
    \draw[->] (H) -- node[auto] {$f = 1$} (T);
    \draw[->] (H) -- node[auto] {$g = 1$} (C);
\end{tikzpicture}
\end{minipage}
\begin{minipage}{.5\textwidth}
\begin{align*}
    S &= N_S \\
    V &= aS + N_V \\
    D &= bV + eC + N_D \\
    T &= cV + fH + N_T \\
    C &= dV + gH + N_C \\
    H &= N_H
\end{align*}
where $N_S, N_V, N_D, N_T, N_C, N_H \overset{\text{iid}}{\sim} \mathcal{N}(0,1)$
\end{minipage}
\caption{Assumed true additive Gaussian noise model (cf.\ Section~\ref{secStimExample}).}\label{figStimulus}
\end{figure*}

\begin{figure*}
\center\
\begin{tikzpicture}
\begin{axis}[
    hide axis,
    clip=false,
    colormap name=divcolor,
    x=.5cm, y=.5cm,
    xmin=1, xmax=23,
    ymin=-4.5, ymax=6.5,
    domain=-7:7,
    point meta min=-2,
    point meta max=2,
    colorbar,
    colorbar horizontal,
    colorbar style={
        xtick={-2, 0, 2},
        xticklabels={$-$, $0$, $+$},
        xticklabel style={
            text width=1em,
            align=center,
            /pgf/number format/.cd,
            fixed
        },
        height=.3cm
    }
    ]

    \brainS{0}{0}{true}{875}{0}{875}{1000}

    \brainS{7}{0}{$\operatorname{enc}(S)$}{875}{0}{875}{1000}

    \brainS{14}{0}{$\operatorname{dec}(V,D,C,T)$}{750}{500}{500}{500}

    \brainS{21}{0}{$\operatorname{dec}(D,C,T)$}{}{375}{375}{625}
\end{axis}
\end{tikzpicture}
\caption{Toy brain maps of the true effects of $S$ in the model described in Figure~\ref{figStimulus} as well as the effects as inferred by three different models (cf.\ Section~\ref{secStimExample}).}\label{figStimulusPlots}
\end{figure*}

\subsection{Response-based paradigm}\label{secRespExample}

Next, assume we conduct a response-based experiment to investigate the neurophysiological causes of cognition (e.\,g.\ assessing subjects' performance in a motor task while recording their brain activity in different regions of interest).
Assume further that the true relationships are governed by the additive Gaussian noise model desribed in Figure~\ref{figResponse}.
The linear effect of each variable onto $P$ is determined by the sum of weight products along each path from that variable to $P$;
e.\,g.\ the effect of $C$ onto $P$ is linear with slope $dg = -1$ (i.\,e.\ increased average activity in $C$ leads to a likewise decreased motor task performance).
The true effects of $T,C,M,D$ onto $P$ are $1,-1,1,0$ respectively.

As in the previous section, Figure~\ref{figResponsePlots} shows the true effects of each variable onto $P$ as well as the outcome one would obtain when running one of the following three analyses of a system admitting the above ground-truth model:
\begin{description}

\item[Analysis 1: $\operatorname{enc}(R)$]\ \\
A mass-univariate encoding analysis where for each variable $X \in \{T,C,M,D\}$ we interpret the weight $\beta_X$ in the model
$$\widehat{X} = \operatorname{enc}(R) = \alpha_X + \beta_X R$$
to reflect the relation between $X$ and $R$.
The true ordinary least squares weights one would obtain in the limit of infinite data are
\begin{align*}
\resizebox{\linewidth}{!}{
$\begin{bmatrix}
    \beta_T \\ \beta_C \\ \beta_M \\ \beta_D
\end{bmatrix}
=
\frac{1}{\var(R)}
\begin{bmatrix}
    \cov(T,R) \\ \cov(C,R) \\ \cov(M,R) \\ \cov(D,R)
\end{bmatrix}
=
\begin{bmatrix}
    \frac{-1}{5} \\ \frac{-4}{5} \\ \frac{4}{5} \\ \frac{-12}{5}
\end{bmatrix}$
}
\end{align*}

\item[Analysis 2: $\operatorname{dec}(T,C,M,D)$]\ \\
A decoding analysis including the variables $T,C,M,D$ where we interpret the weights of the model
$$\widehat{R} = \alpha + \beta T + \gamma C + \delta M + \epsilon D$$
to reflect the effect of each respective variable onto $P$.
The true ordinary least squares weights one would obtain in the limit of infinite data are
\begin{align*}
\begin{bmatrix}
    \beta \\ \gamma \\ \delta \\ \epsilon
\end{bmatrix}
=
\Sigma_{T,C,M,D}^{-1}
\begin{bmatrix}
    \cov(T,R) \\ \cov(C,R) \\ \cov(M,R) \\ \cov(D,R)
\end{bmatrix}
=
\begin{bmatrix}
    0 \\ 0 \\ 1 \\ 0
\end{bmatrix}
\end{align*}

\item[Analysis 3: $\operatorname{dec}(T,C,D)$]\ \\
A decoding analysis as above, this time excluding the variable $M$.
The true ordinary least squares weights one would obtain in the limit of infinite data are
\begin{align*}
\begin{bmatrix}
    \beta \\ \gamma \\ \epsilon
\end{bmatrix}
=
\Sigma_{T,C,D}^{-1}
\begin{bmatrix}
    \cov(T,R) \\ \cov(C,R) \\ \cov(D,R)
\end{bmatrix}
=
\begin{bmatrix}
    \frac{1}{5} \\ \frac{1}{5} \\ \frac{-2}{5}
\end{bmatrix}
\end{align*}

\end{description}

In contrast to the previous example, an encoding analysis does not correctly identify the relations between neurophysiological variables and $P$; e.\,g.\ it turns out that $\beta_T=\frac{-1}{5}$ which may mislead researchers to conclude that $T$ is negatively related to $P$ while indeed it has a positive effect with weight $1$.
The full linear decoding model also leads to incorrect interpretations; e.\,g.\ in this model $\beta$ turns out to be zero suggesting that $T$ is not related to $P$.
Lastly, if we are in the setting where $M$ is unobserved or excluded from the analysis then we arrive at yet another interpretation; e.\,g.\ in this model $\beta = \frac{1}{5}$ systematically underestimates the effect of $T$ onto $P$ and $\gamma=\frac{1}{5}$ suggests a positive effect of $C$ onto $P$ while it indeed has a negative effect.

\begin{figure*}
\begin{minipage}{.5\textwidth}
\begin{tikzpicture}
    \node (T) at(0, 3) {$T${\tiny arget}};
    \node (C) at(3, 3) {$C${\tiny ompensation}};
    \node (M) at(1.5, 1.5) {$M${\tiny otor}};
    \node (P) at(1.5, 0) {$P${\tiny erformance}};
    \node (D) at(4.5, 1.5) {$D${\tiny isturbance}};
    \node (H) at(1.5, 4.5) {$H${\tiny idden}};
    \draw[->] (T) -- node[auto] {$c = 1$} (M);
    \draw[->] (C) -- node[auto] {$d = -1$} (M);
    \draw[->] (M) -- node[auto] {$g = 1$} (P);
    \draw[->] (C) -- node[auto] {$e = 1$} (D);
    \draw[->] (M) -- node[below] {$f = -2$} (D);
    \draw[->] (H) -- node[auto] {$a = 2$} (T);
    \draw[->] (H) -- node[auto] {$b = 3$} (C);
\end{tikzpicture}
\end{minipage}
\begin{minipage}{.5\textwidth}
\begin{align*}
    H &= N_H \\
    T &= aH + N_T \\
    C &= bH + N_C \\
    M &= cT + dC + N_M \\
    D &= eC + fM + N_D \\
    P &= gM + N_P
\end{align*}
where $N_H, N_T, N_C, N_M, N_D, N_P \overset{\text{iid}}{\sim} \mathcal{N}(0,1)$
\end{minipage}
\caption{Assumed true additive Gaussian noise model (cf.\ Section~\ref{secRespExample}).}\label{figResponse}
\end{figure*}

\begin{figure*}
\center\
\begin{tikzpicture}
\begin{axis}[
    hide axis,
    clip=false,
    colormap name=divcolor,
    x=.5cm, y=.5cm,
    xmin=1, xmax=23,
    ymin=-4.5, ymax=6.5,
    domain=-2:2,
    point meta min=-2,
    point meta max=2,
    colorbar,
    colorbar horizontal,
    colorbar style={
        xtick={-2, 0, 2},
        xticklabels={$-$, $0$, $+$},
        xticklabel style={
            text width=1em,
            align=center,
            /pgf/number format/.cd,
            fixed
        },
        height=0.3cm
    }
    ]

    \brainR{0}{0}{true}{900}{100}{900}{500}

    \brainR{7}{0}{$\operatorname{enc}(R)$}{400}{200}{800}{0}

    \brainR{14}{0}{$\operatorname{dec}(T,C,M,D)$}{500}{500}{900}{500}

    \brainR{21}{0}{$\operatorname{dec}(T,C,D)$}{600}{600}{}{300}
\end{axis}
\end{tikzpicture}
\caption{Toy brain maps of the true causes of $P$ in the model described in Figure~\ref{figResponse} as well as the causes as inferred by three different models (cf.\ Section~\ref{secRespExample}).}\label{figResponsePlots}
\end{figure*}
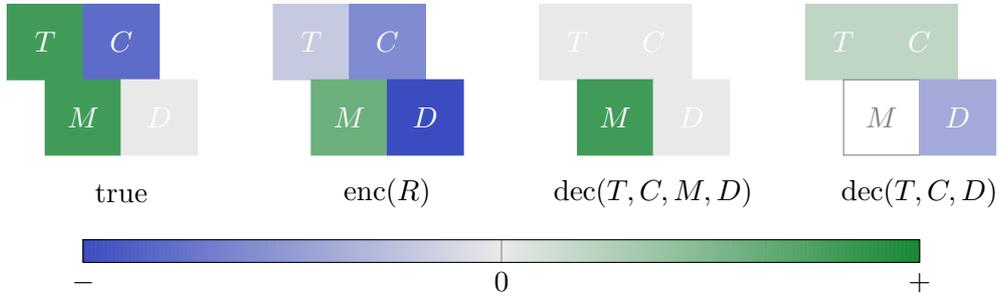

\subsection{Results}

Only the encoding analysis in the stimulus-based setting yields the desired results both qualitatively and quantitatively.
All other analyses in both the stimulus- and response-based setting may lead to incorrect and contradictory interpretations.

\section{Conclusion}

In some particular instances encoding and decoding models may warrant limited partial answers to the questions mentioned at the outset.
However, as above examples demonstrate, the interpretation crucially depends on the experimental paradigm employed, which variables are observable/observed, and which ones enter into the analysis~\citep{waldorp2011effective}.
In general, encoding and decoding models are descriptive models that are fitted to explain observed data and are not designed to answer aforementioned questions.
The cases in which they provide reliable information about effective relationships are somewhat coincidental and scarce~\citep{weichwald2015causal}.

We argue that, if the aim is to answer these questions, we need to advance beyond the encoding versus decoding dichotomy and consider and develop methods specifically tailored to investigate cause-effect relationships (for examples refer to~\citep{chen2015high,grosse2016identification,weichwald2016recovery,weichwald2016merlin}).
We shall not let ourselves be put off with the slogan ``correlation is not causation'' and instead tackle and openly discuss the subtle problems in answering the core neuroimaing questions.
In particular, there has been remarkable progress in very carefully and rigorously researching the required assumptions and theoretical underpinnings of causal inference from experimental data (e.\,g.\ \citep{hoyer2009nonlinear,meinshausen2016methods,mooij2016distinguishing,peters2016causal,scholkopf2016modeling}).
Future research should focus on the right tools for the right questions.

\bibliography{references}

\end{document}